\documentstyle[prl,aps]{revtex}
\begin{document}
\twocolumn[%
\hsize\textwidth\columnwidth\hsize\csname@twocolumnfalse\endcsname
\title{%
\hfill{\normalsize\vbox{\hbox{October 2000} \hbox{DPNU-00-38} }}\\
\vspace{-0.5cm}
\bf Vector Manifestation of the Chiral Symmetry}
\author{{\bf Masayasu Harada} and {\bf Koichi Yamawaki}}
\address{Department of Physics, Nagoya University,
Nagoya, 464-8602, Japan.}
\maketitle

\begin{abstract}
We propose ``vector manifestation (VM)'' of the Wigner realization of
the chiral symmetry in
which the symmetry is restored at the critical point 
by the massless degenerate pion (and its flavor
partners) and rho meson (and its flavor partners) as the chiral partner,
in sharp contrast to the traditional manifestation {\' a} la linear
sigma model where the symmetry is restored by the degenerate pion and
scalar meson.
The application to the chiral phase transition of the large
$N_f$ QCD is performed using the hidden local symmetry Lagrangian.
Combined with the
Wilsonian matching proposed recently, the VM determines the critical
number of massless flavors
$N_f \simeq 5$ without much ambiguity.
\end{abstract}
\vskip1pc]

Chiral phase transition in QCD
is discussed in various contexts such as the large $N_f$ QCD, QCD at
finite temperature and/or density, etc..
In
many situations the traditional linear sigma model-like Wigner
realization for the chiral restoration is assumed.  However, the
Wigner realization does not necessarily require the massless degenerate
pion and scalar meson at the critical point.  
The linear sigma model is merely consistent with the
Wigner realization.  It, therefore, is natural for us to ask the
following question:  Is there a manifestation of the 
Wigner realization other than that of the linear sigma model?
The answer is yes, which we demonstrate in this paper.

In this paper we
propose  ``Vector Manifestation (VM)'' of the chiral symmetry
as a novel manifestation of the Wigner realization
in which the vector meson denoted by $\rho$ ($\rho$ meson and its
flavor partner)
becomes massless at the chiral
phase transition point.  Accordingly, the (longitudinal) $\rho$
becomes the chiral partner of the Nambu-Goldstone (NG) boson denoted
by $\pi$ (pion and its flavor partners).

The essence of VM stems from the new matching 
 of the effective field theory (EFT) with QCD 
(``Wilsonian matching'') recently proposed by
Ref.~\cite{HY:matching} in which bare parameters of the
EFT are determined by matching  
 the current correlators in the EFT
with those obtained by the operator product expansion (OPE) in QCD,
based on the renormalization-group equation (RGE) in the Wilsonian 
sense {\it including the quadratic divergence}~\cite{HY:letter}.
The quadratic divergence was identified with the presence of
a pole of ultraviolet origin at $n=2$ in the dimensional
regularization.
Several physical quantities for $\pi$ and $\rho$ were predicted by 
the Wilsonian matching in the framework of the Hidden Local Symmetry
(HLS)~\cite{BKUYY,BKY} as the EFT, in excellent agreement
 with the experiments for $N_f=3$,
where $N_f$ is the number of 
{\it massless flavors}~\cite{HY:matching}.
This encourages
us to perform the analysis for larger $N_f$ up to near the critical
point based on the Wilsonian matching.

Actually, the chiral symmetry restoration in Wigner realization should
be characterized by the equality of the vector and axialvector current
correlators.
When we approach to the critical point from the broken phase (NG
phase), the axialvector current correlator is still dominated by 
the massless $\pi$ as the NG boson, 
while the vector current correrator is
by the massive $\rho$. 
The crucial ingredient 
of the Wilsonian matching is
the {\it quadratic divergence} which yields 
the quadratic running of (square of) the 
decay constant $F_\pi^2(\mu)$~\cite{HY:letter}, 
where $\mu$ is the renormalization point. It was actually 
shown~\cite{HY:letter} that 
the order parameter $F_\pi(0)$ can become zero for larger $N_f$ 
even when $F_\pi(\Lambda) \ne 0$, 
where $F_\pi(\Lambda)$ is not the order parameter but just a
parameter of the bare Lagrangian defined at the cutoff
$\Lambda$ where the
matching with QCD is made.
Then {\it the $\pi$ contribution to the axialvector current correlator
at $\mu \ne 0 $ persists, $F_\pi(\mu) \ne 0$, \it even at the critical 
point where $F_\pi(0)=0$}. 
Thus the only possibility for this equality to hold at any $\mu \ne 0$ 
is that {\it the $\rho$ contribution to the vector current correlator 
also persists at the critical point in such a way that 
$\rho$ yields a massless pole with the current coupling equal to
that of $\pi$}. Then this restoration, VM, is accompanied by the
degenerate 
massless $\pi$ and (longitudinal) $\rho$ (transverse $\rho$ is
decoupled from 
the current correlator at the critical point, see later discussions).

This is sharply contrasted with
the traditional manifestation of the linear sigma model where the
equality of the current corrrelators is trivially satisfied, since the
axialvector
correlator goes to zero  due to $F_\pi(\mu) \equiv 0$ independently of
$\mu$ (in the absence of the
quadratic divergence), while the vector correlator
has no contribution from the scalar meson and hence is simply zero.
Thus the Wilsonian matching (which leads to $F_\pi(\Lambda) \ne 0$)
excludes the linear sigma model manifestation in QCD.
 
In VM we have degenerate massless $\pi$ and (longitudinal) $\rho$ 
at the phase transition
point,
which are the chiral partners in the
representation of 
$(N_f^2-1\,,\,1) \oplus (1\,,\,N_f^2-1)$  of the chiral 
SU($N_f$)$_{\rm L} \times$SU($N_f$)$_{\rm R}$.
This representation corresponds to $(8\,,\,1)\oplus (1\,,\,8)$ for
$N_f=3$. This is contrasted with the linear sigma
model-like manifestation in which $\pi$ is in the representation
of pure $(3\,,\,3^*)\oplus(3^*\,,\,3)$ together with the scalar meson.
This can be understood in the good-old-day 
saturation scheme of Adler-Weisberger sum
rule for the zero helicity mesons~\cite{WGH}: $\pi$ and the
(longitudinal)
axialvector mesons denoted by $A_1$ ($a_1$ meson and its flavor 
partners) are admixture of $(8\,,\,1) \oplus(1\,,\,8)$ and 
$(3\,,\,3^*)\oplus(3^*\,,\,3)$, since the symmetry is spontaneously
broken:
\begin{eqnarray}
\vert \pi\rangle &=&
\vert (3\,,\,3^*)\oplus (3^*\,,\,3) \rangle \sin\psi
+
\vert(8\,,\,1)\oplus (1\,,\,8)\rangle  \cos\psi
\ ,
\nonumber
\\
\vert A_1\rangle &=&
\vert (3\,,\,3^*)\oplus (3^*\,,\,3) \rangle \cos\psi 
- \vert(8\,,\,1)\oplus (1\,,\,8)\rangle  \sin\psi
\ ,
\nonumber\\
&&\quad
\end{eqnarray}
where the experimental value of the mixing angle $\psi$ is 
given by approximately 
$\psi=\pi/4$~\cite{WGH}.  
 On the other hand, the longitudinal $\rho$
belongs to $(8\,,\,1)\oplus (1\,,\,8)$
and the scalar meson to 
$(3\,,\,3^*)\oplus (3^*\,,\,3)$.
Then the conventional linear sigma model-like manifestation
corresponds to 
the limit $\psi \rightarrow \pi/2$, while the VM
to the limit $\psi\rightarrow 0$ in which case $A_1$ 
goes to a pure $(3\,,\,3^*)\oplus (3^*\,,\,3)$, now degenerate with
the scalar meson in the same representation  $(3\,,\,3^*)\oplus
(3^*\,,\,3)$, but not with $\rho$ in  $(8\,,\,1)\oplus (1\,,\,8)$.

Now we formulate the VM more explicitly.
Let us write the axialvector and vector current correlators
evaluated by the OPE in QCD~\cite{SVZ}:
\begin{eqnarray}
&&
\Pi_A^{\rm(QCD)}(Q^2) = \frac{1}{8\pi^2}
\Biggl[
  - \left( 1 + \frac{\alpha_s}{\pi} \right) \ln \frac{Q^2}{\mu^2}
\nonumber\\
&& \qquad
  + \frac{\pi^2}{3} 
    \frac{
      \left\langle 
        \frac{\alpha_s}{\pi} G_{\mu\nu} G^{\mu\nu}
      \right\rangle
    }{ Q^4 }
  + \frac{\pi^3}{3} \frac{1408}{27}
    \frac{\alpha_s \left\langle \bar{q} q \right\rangle^2}{Q^6}
\Biggr]
\ ,
\nonumber\\
&&
\Pi_V^{\rm(QCD)}(Q^2) = \frac{1}{8\pi^2}
\Biggl[
  - \left( 1 + \frac{\alpha_s}{\pi} \right) \ln \frac{Q^2}{\mu^2}
\nonumber\\
&& \qquad
  + \frac{\pi^2}{3} 
    \frac{
      \left\langle 
        \frac{\alpha_s}{\pi} G_{\mu\nu} G^{\mu\nu}
      \right\rangle
    }{ Q^4 }
  - \frac{\pi^3}{3} \frac{896}{27}
    \frac{\alpha_s \left\langle \bar{q} q \right\rangle^2}{Q^6}
\Biggr]
\ ,
\label{Pi A V OPE}
\end{eqnarray}
where $\mu$ is the renormalization scale of QCD, $Q$ the Euclidean
momentum carried by the current, and we neglected 
${\cal O}\left(1/Q^8\right)$ terms.
These expressions are valid in high energy where the QCD
coupling $\alpha_s$ is small.

Next we consider the expression of the current correlators in the EFT
which is valid in the low energy below the matching scale $\Lambda$.
As an EFT to describe the VM we need a model having both $\pi$ and
$\rho$ fields.
Here we use the HLS model~\cite{BKUYY,BKY} which includes
$\pi$ and $\rho$ consisitently with the chiral symmetry and
actually reproduces experiments nicely
through the Wilsonian matching~\cite{HY:matching}.
The axialvector and vector current correlators
in the HLS are well described by the tree contributions with including
${\cal O}(p^4)$ terms when the momentum is around the matching scale
$\Lambda$~\cite{HY:matching}:
\begin{eqnarray}
\Pi_A^{\rm(HLS)}(Q^2) &=&
\frac{F_\pi^2(\Lambda)}{Q^2} - 2 z_2(\Lambda)
\ ,
\nonumber\\
\Pi_V^{\rm(HLS)}(Q^2) &=&
\frac{
  F_\sigma^2(\Lambda)
  \left[ 1 - 2 g^2(\Lambda) z_3(\Lambda) \right]
}{
  M_v^2(\Lambda) + Q^2
} 
- 2 z_1(\Lambda)
\ ,
\label{Pi A V HLS}
\end{eqnarray}
where $g(\Lambda)$ is the bare HLS gauge coupling,
$F_\sigma^2(\Lambda) = a(\Lambda) F_\pi^2(\Lambda)$ is the bare decay
constant of the would-be NG boson $\sigma$ (not to be confused with
the scalar meson in the linear sigma model) 
absorbed into the HLS gauge boson, and 
$M_v^2(\Lambda) \equiv g^2(\Lambda) F_\sigma^2(\Lambda)$ is the bare
HLS gauge boson mass.
In Ref.~\cite{HY:matching} these correlators are matched with those in
Eq.~(\ref{Pi A V OPE}) up to the second
derivative in terms of $Q^2$ for $Q^2=\Lambda^2$.  The resultant
Wilsonian matching condition relevant to the present analysis is
given by~\cite{HY:matching}
\begin{eqnarray}
&&
\frac{F_\pi^2(\Lambda)}{\Lambda^2} 
= \frac{1}{8\pi^2}
\Biggl[
  1 + \frac{\alpha_s}{\pi}
\nonumber\\
&& \qquad\quad
  + \frac{2\pi^2}{3} 
    \frac{
      \left\langle 
        \frac{\alpha_s}{\pi} G_{\mu\nu} G^{\mu\nu}
      \right\rangle
    }{ \Lambda^4 }
  + \pi^3\, \frac{1408}{27}
    \frac{\alpha_s \left\langle \bar{q} q \right\rangle^2}%
    {\Lambda^6}
\Biggr]
\ .
\label{match A}
\end{eqnarray}

Let us now obtain constraints on the bare parameters of the HLS in the
VM through the Wilsonian matching.
At the critical point
the quark condensate
$\left\langle \bar{q} q \right\rangle$ vanishes, while
the gluonic condensate 
$\left\langle 
\frac{\alpha_s}{\pi} G_{\mu\nu} G^{\mu\nu}
\right\rangle$
is independent of the renormalization point of
QCD and hence  it is expected that it
does not vanish.
Then the right-hand-side (RHS) of Eq.~(\ref{match A}) is non-zero,
implying that {\it $F_\pi^2(\Lambda)$ is non-zero even at the critical
point}.

Then how do we know by the bare parameters defined at $\Lambda$
whether or not the chiral symmetry is restored ?
As we discussed before, a clue comes from the fact that 
$\Pi_A^{\rm (QCD)}$ 
and $\Pi_V^{\rm (QCD)}$ in Eq.~(\ref{Pi A V OPE})
agree with each other for any value of $Q^2$ when the chiral symmetry
is restored with $\left\langle \bar{q} q \right\rangle = 0$.
Thus, we require that
$\Pi_A^{\rm (HLS)}$ and $\Pi_V^{\rm (HLS)}$ in Eq.~(\ref{Pi A V HLS})
agree with each other for {\it any value of $Q^2$}.
This agreement is satisfied only if the following conditions are met:
\begin{eqnarray}
&& g(\Lambda) \rightarrow 0 \ , 
\quad a(\Lambda) \rightarrow 1 \ , \nonumber\\
&& z_1(\Lambda) - z_2(\Lambda) \rightarrow 0 \ .
\label{vector conditions}
\end{eqnarray}
This is nothing but the VM of the chiral symmetry in terms of the 
HLS parameters.
Note that $a(\Lambda)\simeq 1$ is satisfied in
QCD already for $N_f=3$ in the broken phase~\cite{HY:matching}.
The first two in Eq.~(\ref{vector conditions})
are the values in the Georgi's vector limit~\cite{Georgi}, 
which was simply assumed in Ref.~\cite{HY:letter} to be a 
consistent way to incorporate the
chiral phase transition of the large $N_f$ QCD
into the HLS. Thanks to the Wilsonian matching it is
now clear that {\it Eq.~(\ref{vector conditions}) is 
the precise HLS expression of the Wigner realization in QCD}.

The VM in the HLS
is similar to the Georgi's ``vector
realization''~\cite{Georgi}, 
but is different in an essential way:
The ``vector realization'' is claimed to be a different realization 
than either the Wigner or 
NG realizations in such a way that 
the NG boson
does exist ($F_\pi(0)\ne0$) while the chiral symmetry is still unbroken.
On the contrary, our  VM is precisely the Wigner
realization having $F_\pi(0) =0$.
Technically, the bare HLS Lagrangian in the VM
coincides with the parameter choice of the Georgi's ``vector
realization'';  $g(\Lambda)=0$, $a(\Lambda)=1$ and
$F_\pi(\Lambda)\neq 0$.
However, an essential difference comes from the Wilsonian RGE's whose 
quadratic divergence  leads to the Wigner realization with
$F_\pi(0)=0$ at the low-energy limit (on-shell of NG bosons).
On the other hand, the ``vector realization''
lacking the quadratic divergence leads to $F_\pi(0)=F_\pi(\Lambda)\ne 0$.
In contrast to the Georgi's ``vector realization'', the VM
in the Wigner realization is consistent with the
chiral Ward-Takahashi identity~\cite{Y,HY:prep}.

We now examine the chiral symmetry restoration in the 
large $N_f$ QCD
($N_f < \frac{11}{2}N_c$) which was implied by the fact 
that the coupling at the infrared fixed
point becomes very small~\cite{BZ}.
Such a restoration was indeed observed by various methods like 
lattice simulation~\cite{lattice}, 
ladder 
Schwinger-Dyson (SD) equation~\cite{ATW,MY},
dispersion
relation~\cite{OZ}, instanton calculus~\cite{VS}, etc..
As pointed out in Ref.~\cite{MY} the phase transition for large $N_f$
QCD may be characterized by the ``conformal phase transition''.
In such a case the
Ginzburg-Landau effective theory (linear sigma model-like
manifestation) simply breaks down.
The VM may be a manifestation of the chiral
symmetry restoration consistent with the ``conformal phase
transition''.

The chiral restoration in terms of HLS 
was obtained in Ref.~\cite{HY:letter}
without VM and Wilsonian matching.
What was shown in Ref.~\cite{HY:letter} is that the RGE for
$F_\pi^2$ including the quadratic divergence
reduces the 
value of $F_\pi^2(0)$ (the pole residue of the massless pion
pole) from
$F_\pi^2(\Lambda)$
(the bare parameter of the HLS Lagrangian defined at a cutoff scale
$\Lambda$) 
in such a way that the larger $N_f$,  the smaller the value of
$F_\pi(0)$ is. 
It eventually goes to zero, the chiral restoration, 
at a certain critical number of $N_f$.

Now in
the VM,  the bare parameters are characterized by
the vector limit $g(\Lambda)=0$ and $a(\Lambda)=1$ 
[see Eq.~(\ref{vector conditions})] which is actually the fixed point
of RGE~\cite{HY:letter}.
Then the VM justifies the previous
derivation of the RGE for $F_\pi$ in the vector limit 
which relates the order parameter with the bare parameter 
as~\cite{HY:letter}:
\begin{equation}
\frac{F_\pi^2(m_\rho=0;N_f)}{\Lambda_f^2} = 
\frac{F_\pi^2\left(\Lambda_f;N_f\right)}{\Lambda_f^2}
- \frac{N_f}{2(4\pi)^2} \ ,
\label{RGE for fpi2 at vector limit}
\end{equation}
where we expressed the matching scale by $\Lambda_f\equiv\Lambda(N_f)$
since it generally depends on $N_f$.
It should be noticed that this equation holds only at the critical
flavor $N_f^{\rm cr}$, and the left-hand-side vanishes at the critical
point.

The value of $N_f^{\rm cr}$ is determined in terms of
the parameters in the OPE by combining 
Eq.~(\ref{RGE for fpi2 at vector limit}) and Eq.~(\ref{match A}) with
taking $\left\langle \bar{q} q \right\rangle=0$.
The resultant expression is given by
\begin{equation}
N_f^{\rm cr} = 4 
\left[
  1 + \frac{\alpha_s}{\pi}
  + \frac{2\pi^2}{3} 
    \frac{
      \left\langle 
        \frac{\alpha_s}{\pi} G_{\mu\nu} G^{\mu\nu}
      \right\rangle
    }{ \Lambda_f^4 }
\right]
\ .
\end{equation}
Here we estimate this by using the parameters for
$N_f=3$~\cite{HY:matching}:
$(\Lambda_3\,,\,\Lambda_{\rm QCD})=(1.2\,,\,0.35)$\,GeV,
$\alpha_s=0.56$ and 
$\left\langle \frac{\alpha_s}{\pi} G_{\mu\nu} G^{\mu\nu}
\right\rangle=0.012\,\mbox{GeV}^4$.
The result is given by
\begin{equation}
N_f^{\rm cr} \simeq 4.9 \ .
\label{crit val}
\end{equation}
The precise estimation of this will be done by determining the
$N_f$-dependences of the QCD coupling $\alpha_s$ and $\Lambda_f$ in
the forthcoming paper~\cite{HY:prep}.
Here we just quote the result $N_f^{\rm cr}\simeq4.8$,
which is consistent with the above estimate.

To study the critical behaviors of the parameters
when approaching to the critical point, we need to know
how the bare parameters
$g(\Lambda_f;N_f)$ and $a(\Lambda_f;N_f)$ approach to 
the vector limit Eq.~(\ref{vector conditions}).
Comparing the difference between vector and axialvector correlators 
in Eq.~(\ref{Pi A V OPE}) with that in Eq.~(\ref{Pi A V HLS}), 
we know that the critical behavior of $g^2(\Lambda_f;N_f)$
is given as $g^2(\Lambda_f;N_f) \sim \left\langle \bar{q} q
\right\rangle^2$. 
Since we do not know the scaling of $\left\langle \bar{q} q 
\right\rangle$ except for the ladder SD
approach\cite{foot:scaling},
we here tentatively
adopt the following ansatz on the behavior of the HLS gauge coupling
approaching to zero:
\begin{equation}
g^2(\Lambda_f;N_f) 
= 
\bar{g}^2 \epsilon^q \ , \quad
\epsilon \equiv
1/N_f - 1/N_f^{\rm cr} \ ,
\label{initial value of g}
\end{equation}
where $\bar{g}$ is independent of $N_f$.~\cite{foot:couple}
Moreover,
we fix $a(\Lambda_f;N_f) = 1$ even off the critical point, since
the Wilsonian matching conditions
with the physical inputs $F_\pi(0)=88$\,MeV and $m_\rho=770$\,MeV
 leads to $a(\Lambda)\simeq 1$ already  for
$N_f=3$~\cite{HY:matching}.
The RGE's for $F_\pi^2$ and $g^2$ are analytically solvable for
$a=1$.
A careful analysis~\cite{HY:prep} leads to that $q$ in 
Eq.~(\ref{initial value of g}) must satisfy $q\ge1$ for the
consistency.
The resultant critical behaviors of 
the order parameter and the mass of $\rho$ are given by
\begin{eqnarray}
&&
F_\pi^2(0;N_f)/\Lambda_f^2 \sim \epsilon \rightarrow 0
\ ,
\nonumber\\
&&
m_\rho^2(N_f)/\Lambda_f^2 \sim
\epsilon^{ 1 + q } \rightarrow 0 \ ,
\label{critical of fpi and mrho}
\end{eqnarray}
which shows that {\it $m_\rho$ approaches to zero faster than
$F_\pi$}.  This is a salient feature of the VM~\cite{foot:crit}.

Let us consider the critical behaviors of the physical quantities
listed in Ref.~\cite{HY:matching}.
The $\rho$--$\gamma$ mixing strength $g_\rho$ and the
$\rho$-$\pi$-$\pi$ coupling constant $g_{\rho\pi\pi}$ go to zero as
\begin{eqnarray}
&& g_\rho(m_\rho) = g(m_\rho) F_\pi^2(m_\rho)
\sim \epsilon^{1+q/2} \rightarrow 0 \ ,
\nonumber\\
&& g_{\rho\pi\pi}(m_\rho,0,0) = \frac{g(m_\rho)}{2}
\frac{F_\pi^2(m_\rho)}{F_\pi^2(0)} \sim \epsilon^{q/2} 
\rightarrow 0 \ ,
\label{eq:grho}
\end{eqnarray}
where $a(\Lambda)=a(m_\rho)=1$ was used.
As discussed in Ref.~\cite{HY:matching}, the KSRF (I) relation for the
low-energy quantities 
$g_\rho(0) = 2 g_{\rho\pi\pi}^2(0,0,0) F_\pi^2(0)$
holds as a low energy theorem of the HLS~\cite{BKY,HY,LET:2}
for any $N_f$.
The relation for on-shell quantities is violated by about 10\% for
$N_f=3$~\cite{HY:matching}.
As $N_f$ goes to $N_f^{\rm cr}$, $g_\rho(m_\rho)$ and 
$g_{\rho\pi\pi}(m_\rho,0,0)$ approach to
$g_\rho(0)$ and $g_{\rho\pi\pi}(0,0,0)$, respectively,
and hence the on-shell KSRF (I) relation becomes more
accurate for larger $N_f$.
On the other hand, the (on-shell) KSRF (II) relation $m_\rho^2 = 2
g_{\rho\pi\pi}^2(m_\rho,0,0) F_\pi^2(0)$ becomes less accurate.  
Near the critical flavor it reads as $m_\rho^2 = 4
g_{\rho\pi\pi}^2(m_\rho,0,0) F_\pi^2(0) \rightarrow 0$.

Several comments are in oder:

In the VM both the axialvector and vector
current correlators in Eq.(\ref{Pi A V HLS}) take the form of
$F_\pi^2(\Lambda)/Q^2 - 2 z_2(\Lambda)$.
For the axialvector current
correlator, the first term $F_\pi^2(\Lambda)/Q^2$ comes from
the $\pi$-exchange contribution, while 
for the vector current correlator it can be easily understood as the
$\sigma$ (would-be NG boson absorbed into $\rho$)-exchange
contribution in the $R_\xi$-like gauge.
Thus only the longitudinal
$\rho$ couples to the vector current, and 
{\it the transverse $\rho$ with the helicity
$\pm 1$, which belongs to the representation $(N_f,N_f^*)\oplus
(N_f^*,N_f)$, 
is decoupled from it}.
This can be also seen in the unitary gauge.~\cite{HY:prep}

The parameters $L_{10}(m_\rho)$ and $L_9(m_\rho)$ defined in
Ref.~\cite{HY:matching} diverge as
$N_f$ approaches to $N_f^{\rm cr}$.
However, we should note that, even for $N_f=3$, both $L_{10}(\mu)$ and
$L_9(\mu)$ have the infrared logarithmic divergences
when we take $\mu\rightarrow 0$ in
the running obtained by the chiral perturbation theory~\cite{GL}.
Thus we need more careful treatment of these quantities for large
$N_f$.  This is beyond the scope of this paper.

{\it The $A_1$ in the VM is resolved 
and/or decoupled from the axialvector current near the critical
flavor}
since there is no contribution in the vector current
correlator to be matched with the axialvector current
correlator.  As to the scalar meson~\cite{HSS}, although the mass
is smaller than the matching scale adopted in
Ref.~\cite{HY:matching} for $N_f=3$~\cite{foot:scalar}, 
we expect that 
{\it the scalar meson is also resolved and/or decoupled near
the chiral phase transition point}, since it is
in the $(N_f,N_f^*)\oplus (N_f^*,N_f)$ representation together
with the $A_1$ in the VM.

In this paper we applied the VM to the chiral
restoration in the large $N_f$ QCD.
It may be checked by the lattice
simulation:  The vanishing ratio $m_\rho/F_\pi(0)$ is a clear
indication of the VM.

The VM  may be applied to other chiral phase transitions such as the
one at finite temperature and/or density.  In such a case, the
position of the $\rho$ peak of the dilepton spectrum would move to the
lower energy region in accord with the picture shown in
Ref.~\cite{BR}, and Eqs.~(\ref{critical of fpi and mrho}) and
(\ref{eq:grho}) would further imply smaller $\rho$ width 
($\Gamma/m_\rho \sim g_{\rho\pi\pi}^2 \sim \epsilon^{q}$) and
larger peak value  ($\Gamma_{ee}\Gamma_{\pi\pi}/\Gamma^2
\sim g_\rho^2/(g_{\rho\pi\pi}^2 m_\rho^4)
\sim 1/\epsilon^{2q}$) near the critical point.  If it is really the
case, these would  be clear signals of VM tested in the future
experiments.

The VM
studied in this paper may be applied to the models for the composite
$W$ and $Z$.  Our analysis shows that the mass of the
composite vector boson approaches to zero
faster than the order parameter,
which is fixed to the electroweak symmetry breaking scale,
near the critical point. The VM may also be applied to the technicolor 
with light techni-$\rho$.

We would like to thank Howard Georgi for his enlightening and critical
discussions.
K.Y. thanks Howard Georgi for hospitality during the stay at Harvard
where a part of this work was done.
This work is supported in part by Grant-in-Aid for Scientific Research
(B)\#11695030 (K.Y.), (A)\#12014206 (K.Y.) and (A)\#12740144 (M.H.),
and by the Oversea Research Scholar Program of the Ministry of
Education, Science, Sports and Culture (K.Y.).


\begin{thebibliography}{99}
\bibitem{foot:scaling}
If the scaling of $\left\langle \bar{q} q \right\rangle$ is 
of the ``conformal phase transition''\cite{MY} as was suggested by the
SD approach~\cite{ATW,MY}, $g^2(\Lambda_f;N_f)$ has the 
essential-singularity scaling.
\bibitem{foot:couple}
For $q=1$ this behavior may be justified by a large $N_f$
argument. (See Ref.~\cite{HY:letter}.)
\bibitem{foot:crit}
Even if we take the ``conformal phase transition''-type 
for $g^2(\Lambda_f;N_f)$
instead of Eq.~(\ref{initial value of g}), we still have this
feature.
Actually, we obtain the same power
behavior for $F_\pi^2(0;N_f)$ as in 
Eq.~(\ref{critical of fpi and mrho}), because we use the one-loop
RGE's.
\bibitem{foot:scalar}
The scalar meson
does not couple to the axialvector and vector currents, anyway.
\end{thebibliography}
\end{document}